\documentclass[prl,showpacs,amsmath,amssymb,twocolumn]{revtex4}

\usepackage[dvips]{epsfig}
\usepackage[dvips]{graphicx}
\usepackage{dcolumn}
\usepackage{bm}

\graphicspath{{converted_graphics/}}
\begin{document}

\title{Multiscale theory of valley splitting}

\author{Sucismita Chutia}
\author{S. N. Coppersmith}
\author{Mark Friesen}
\affiliation{Department of Physics, University of
Wisconsin-Madison, Wisconsin 53706, USA} 

\begin{abstract}
The coupling between $z$ valleys in the conduction band of a Si quantum 
well arises from phenomena occurring within several atoms from the interface, 
thus ruling out a theoretical description based on pure effective mass theory.  
However, the complexity and size of a realistic device precludes an analytical 
atomistic description.  Here, we develop a fully analytical multiscale theory 
of valley coupling, by combining effective mass and tight binding approaches.  
The results are of particular interest for silicon qubits and quantum devices, 
but also provide insight for GaAs quantum wells.
\end{abstract}

\pacs{73.21.Fg,73.20.-r,78.67.De,81.05.Cy}

\maketitle

The two-dimensional electron gas formed at a silicon heterointerface underpins the 
modern electronics industry.  But in spite of its ubiquity, the silicon 
interface exhibits phenomena that are not fully understood, and cannot
be explained by the conventional effective mass theory.  In the emerging field of 
nanoelectronics, quantum degrees of freedom like spin form the basis
for novel technologies, such as spintronics \cite{zutic04} and quantum computation 
\cite{loss98,kane98}.  An alternative degree of freedom is associated with the
low-lying features in the conduction band structure, known as valleys, for Si and
other indirect gap
semiconductors \cite{ando82}.  The degeneracy of the low-lying valley states is 2-fold 
for both Si[001]/SiO$_2$ and Si[001]/SiGe interfaces, 
due to mass anisotropy and strain effects, respectively.  For quantum devices,
where valley physics has been studied most extensively \cite{goswami07},  
the valley states may either be ``frozen out," in favor of the spin
degree of freedom \cite{friesen03}, or utilized as qubits \cite{smelyanskiy05}.
In either case, a complete understanding of the valley physics is essential.

One main concern is the origin and the magnitude of valley coupling, 
which lifts the valley degeneracy.  The two approaches previously
applied to this problem involve continuum theories like effective mass (EM) 
\cite{kohn,ohkawa77,sham79,friesen05,ahn05,friesen07}, and atomistic theories like tight
binding (TB) \cite{boykin04,boykin04b}.  While the former provides intuition because of 
its analytical nature, it cannot fully account for the fundamentally discrete and atomistic 
nature of the valley coupling.  On the other hand, atomistic theories give an accurate 
description of the valley splitting from first principles, but they cannot 
provide analytical results, except in the simplest geometries \cite{boykin04}.
In this letter, we bridge the gap between microscopic and macroscopic theories
through a multiscale technique, and we provide theoretical justification for an
intuitive extended EM theory.  We show how atomic scale corrections 
near an interface lead to significant improvements in the EM description, 
even for direct gap semiconductors like GaAs.  

It is well known that the conventional EM theory of electron
confinement in a semiconductor crystal breaks down near a sharp confining
potential, like a quantum 
well \cite{ando82}.  For direct gap materials, the resulting errors
are typically small and may be treated perturbatively in 
the long-wavelength EM theory \cite{burt92}.
For indirect gap materials like Si, the atomic scale physics of the interface,
which is absent from the EM theory, is also responsible for 
valley splitting, corresponding to an energy scale ($\sim 1$~meV) that is
comparable with other energies of interest for quantum devices.  
Perturbative treatments should therefore be undertaken with caution.
Here, we describe a multiscale approach 
that efficiently captures the atomic scale corrections by means of an effective
interface potential, $\Lambda$.  The approach is fully analytical, and 
can be applied to general quantum well geometries.

For concreteness, we consider the case of a SiGe/Si/SiGe symmetric square well.
The EM wavefunction is written as \cite{kohn}
\begin{equation}
\Psi({\bm r})=\sum_{n=\pm z} \alpha_ne^{ik_nz} u_{k_n}({\bm r})F(z) ,
\label{eq:kohn}
\end{equation}
where $\alpha_n$ are the valley composition factors (of no importance here),
$k_{\pm z}=\pm k_0$ are the positions of the valley minima in
the Brillouin zone, $u_{k_n}({\bm r})$ are periodic Bloch functions, and
$F(z)$ is the long wavelength envelope.  Note that we have only included 
contributions from the two $z$ valleys, as appropriate for [001] strained
quantum wells.  The short wavelength physics is
contained in the fast phase oscillations and the Bloch functions.  Note that the silicon
valley minima are located near the Brillouin zone
boundaries, with \cite{boykin04} $k_0=0.82 (2\pi /a)$ for a Si cubic unit cell
of width $a=5.43$~\AA, consisting of four atomic planes along $[001]$.

We have previously argued that the atomic scale physics of valley coupling 
can be incorporated 
into the long wavelength theory by means of a $\delta$-function potential 
at the quantum well interface \cite{friesen05}. 
Similar arguments have also been put forth in Refs.~\cite{foreman05,nestoklon06},
leading to a coupled set of envelope equations of the form 
\cite{friesen07,fritzsche,twose}
\begin{eqnarray} &&
\sum_{n=\pm z} \alpha_ne^{ik_nz} \biggl[ -\frac{\hbar^2}{2m_l}
\frac{\partial^2}{\partial z^2} +V(z)
\label{eq:envelope} \\ \nonumber && \hspace{1in}
+\sum_m  \Lambda \delta (z-z_m)-E\biggr] F(z)=0 .
\end{eqnarray}
Here, $m_l=0.916m_0$ is the longitudinal effective mass, $V(z)$ is the
vertical confinement from the conduction band offset of
$V_0$, and $z_m=\pm L/2$ are the two quantum well interfaces.  The $ \Lambda $ term
couples the $z$ valleys.  Since $V(z)$ is 
constant away from the interface, the solutions for the ground state envelope 
function are given by
\begin{equation}
F(z) = \left\{
\begin{array}{ll}
A \cos (qz) & (|z|<L/2) \\
B\, e^{-pz} & (|z|\geq L/2)
\end{array} \right. . \label{eq:EMF}
\end{equation}
Conventionally, the unknown parameters in Eq.~(\ref{eq:EMF}) are determined 
by matching the envelope function and its first derivative on either side of 
the interface \cite{daviesbook}.  However, the latter matching condition must 
be modified in the presence of a $\delta$-function interface potential.  
Integrating Eq.~(\ref{eq:envelope}) over an infinitesimal
range about the interface, we obtain the new matching condition,
\begin{equation}
 \Lambda = \frac{\hbar^2}{2m_l}\frac{F'_+(L/2)-F_-'(L/2)}{F(L/2)} ,
\label{eq:BC}
\end{equation}
where $F'_+$ ($F'_-$) correspond to right-hand (left-hand) derivatives.
Note that a similar discontinuity in $F'(z)$ also occurs at any heterojunction
with two different effective masses \cite{daviesbook}.  This effect is unrelated 
to valley coupling, and we ignore it below.  Indeed, for silicon-rich 
SiGe materials, the mass variations are small and inconsequential for our 
main results.

We now construct a multiscale theory for $\Lambda$.  To begin, we note that 
the conventional EM theory remains valid and accurate, except within
about one atom distance from the quantum well interface.  In the vicinity of the
interface, the EM theory should be replaced by an atomistic one.  The
simplest TB theory that can describe
valley coupling was derived in Ref.~\cite{boykin04}.  The model involves two bands,
with nearest and next-nearest neighbor tunneling parameters, $t_1$ and 
$t_2$, respectively.  An additional onsite parameter describes the confinement
potential $V(z)$ of the quantum well.  Although these parameters may vary with position, 
depending on the alloy composition, the variations are small for SiGe and we 
ignore them here.  The tight-binding coupling parameters are then given by \cite{boykin04}
\begin{equation}
t_2\sin^2(k_0a/4)=2\hbar^2/m_la^2, \hspace{0.15in}
t_1=4t_2\cos (k_0a/4) .
\end{equation}
Near the top interface, the TB Hamiltonian is given by
\begin{equation}
H=\begin{pmatrix}
\ddots &&&&&&&&& \\
&t_2&t_1&\bm{0}&t_1&t_2&0&0&0& \\
&0&t_2&t_1&\bm{0}&t_1&t_2&0&0& \\
&0&0&t_2&t_1&\bm{V_0}&t_1&t_2&0& \\
&0&0&0&t_2&t_1&\bm{V_0}&t_1&t_2& \\
&&&&&&&&&\ddots
\end{pmatrix} . \label{eq:TB}
\end{equation}
(Diagonal elements are highlighted in bold, for clarity.)
The eigenstates of $H$ correspond to vectors of TB coefficients 
$(\dots,C_{-2},C_{-1},C_0,C_1,\dots)$.  
Here, we consider a quantum well of size $L=a(N+1)/2$, containing $(2N+1)$ atoms,
and centered at atomic position $N=0$. 

\begin{figure}[t] 
  \centering
  \includegraphics[bb=24 442 345 754,width=2.3in,height=2.24in,keepaspectratio]{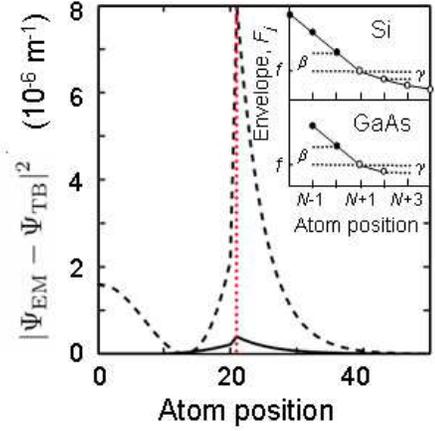}
  \caption{(Color online)
Differences between the effective mass (EM) and tight binding (TB) wavefunction
solutions are shown for a GaAs quantum well.  (Only the right half of the well is shown.)
The dashed line shows the conventional EM theory, while the solid line includes multiscale 
corrections.  The vertical dotted line marks the quantum well boundary.
Inset:  Multiscale ansatz for envelope functions near the interface.  For Si, 7 atoms  
are needed.  For GaAs only 4 atoms are used.  Open circles correspond to sites 
in the barrier region.}
  \label{fig:GaAsfig}
\end{figure}

To implement a multiscale theory, we require that the TB eigenstates
match the EM solutions away from the interface.  From
Eq.~(\ref{eq:kohn}), the two lowest-energy TB wavefunctions can be expressed as
\begin{eqnarray}
C_j &=& (-1)^j \sqrt{2} \cos (jk_0a/4) F_j , \label{eq:TBF} \\
S_j &=& (-1)^j \sqrt{2} \sin (jk_0a/4) F_j , \label{eq:TBFS}
\end{eqnarray}
where the actual ground state depends on the width of the quantum well \cite{friesen07}.
Here, $(-1)^j/\sqrt{2}$ corresponds to the Bloch function for the lowest
band of the two-band model \cite{friesen07}, while the cosine and sine functions
describe the exponential phase factors in Eq.~(\ref{eq:kohn}), corresponding to
the cases $(\alpha_{+z},\alpha_{-z})=(1,\pm 1)/\sqrt{2}$, respectively.
The envelope coefficient $F_j=F(ja/4)$ is determined from the correspondence with 
Eq.~(\ref{eq:EMF}).  The physics of valley coupling
is captured by the sudden change of slope (\textit{i.e.}, the kink) in the envelope
at the interface.  To facilitate calculations, we consider the model parameters
shown in the inset of
Fig.~1.  Immediately adjacent to the interface, the envelope function exhibits 
a change of slope, with $F'_-= (4/a)\beta$ and $F'_+= (4/a)\gamma$
on either side of the kink, and amplitude $F_{N+1}=f$ right at the interface.  Thus,
$F_N=f+\beta$, $F_{N+2}=f-\gamma$, and so on.  

Before solving the multiscale theory, it is illuminating to gauge the accuracy
of the EM ansatz of Eqs.~(\ref{eq:TBF})-(\ref{eq:TBFS}).  This is 
accomplished in Fig.~2, using the two lowest-energy TB eigenstates to infer the
envelope function from the relation $2F_j^2=C_j^2+S_j^2$.  
The result exhibits residual short wavelength structure, 
arising from the fact that the exact $k$ values of the fast oscillations of the two 
lowest eigenstates are nearly (but not quite) identical \cite{boykin04b}.
This is a signature of the incomplete separation of
the long and short wavelength physics, and it places a fundamental limit on our 
ability to match the TB and EM theories.  In the present work, the spurious
``jitter" evident in Fig.~2 leads to errors in the evaluation of the kink.

The multiscale theory involves just two equations from the full TB Hamiltonian, 
and we can choose which equations to use.  The spurious jitter in Fig.~2, could be 
mitigated by an averaging procedure.  Indeed, in this way, we obtain
excellent agreement with previous estimates of the valley splitting \cite{friesen07}.  
However, such techniques detract
from the simplicity of the multiscale approach.  Here, we take a different tack,
noting that the alternating behavior of the TB coefficients in Fig.~2 can be 
partially mitigated simply by using alternating
TB equations.  We consider the following equations centered
symmetrically around the interface:  
\begin{eqnarray}
t_2C_{N-2}+t_1C_{N-1}+t_1C_{N+1}+t_2C_{N+2} &=& E_\text{TB}C_N , \nonumber \\
t_2C_{N}+t_1C_{N+1}+V_0C_{N+2} \hspace{0.75in} && \label{eq:system} \\
+ t_1C_{N+3}+t_2C_{N+4} &=& E_\text{TB}C_{N+2} . \nonumber 
\end{eqnarray}
Note that either the cosine (\ref{eq:TBF}) or sine (\ref{eq:TBFS}) functions
can be used here.

\begin{figure}[t] 
  \centering
  \includegraphics[bb=34 395 428 758,width=2.3in,height=2.12in,keepaspectratio]{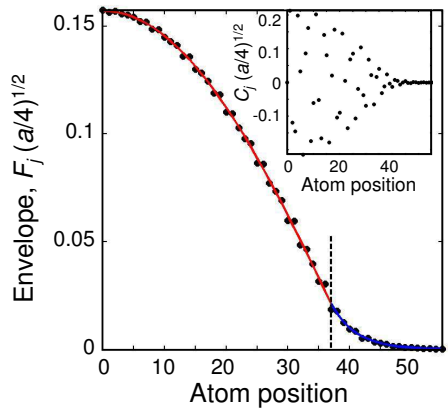}
  \caption{(Color online)  Dimensionless envelope for the ground state wavefunction 
in a 10 nm Si/SiGe quantum well.  (Only the right half of the quantum well is shown.)  
The vertical dashed line marks the quantum well boundary.  The discrete points are 
obtained from TB theory, as described in the text.  
Spurious short wavelength structure is cause by limitations in the EM theory.
Inside (outside) the quantum well, the blue
(red) solid lines are fits to Eq.~(\ref{eq:EMF}).
Inset:  the full, dimensionless TB wavefunction, including the fast oscillations.}
  \label{fig:wv}
\end{figure}

It is necessary to take into account the curvature of the envelope functions in
system~(\ref{eq:system}), in order to avoid unphysical solutions.  
Near the interface, the cosine envelope in Eq.~(\ref{eq:EMF}) has a vanishing 
second derivative, but the exponential envelope does not.
To leading order, we can express the latter in terms of 
parameters $f$ and $\gamma$ as follows:
$F_{N+2}=f-\gamma$, $F_{N+3}=f-2\gamma +\gamma^2/f$,
and $F_{N+4}=f-3\gamma +3\gamma^2/f$.  Evaluating system~(\ref{eq:system}),
we now obtain
\begin{eqnarray}
\frac{F'_+(L/2)-F_-'(L/2)}{F(L/2)} &=& \frac{(\beta -\gamma)}{f}\frac{4}{a} \\
&\simeq& \frac{\gamma^2}{f^2}=\left(\frac{pa}{4}\right)^2 
\simeq \frac{mV_0a^2}{8\hbar^2}, \nonumber
\end{eqnarray}
where we have dropped higher order terms in the small parameter $a/L$.
The anticipated linear dependence of $\Lambda$ on $V_0$ \cite{friesen07} emerges
from Eq.~(\ref{eq:BC}):
\begin{equation}
 \Lambda \simeq V_0a/4=(1.36 \times 10^{-10})V_0 , \label{eq:final}
\end{equation}
where $\Lambda$ is in units of eVm when $V_0$ is in eV.

Eq.~(\ref{eq:final}) is our main result, obtained through a multiscale analysis
of the kink of the envelope function.  We can compare this with the apparent kink
obtained by fitting Eq.~(\ref{eq:EMF}) to the full TB envelope function, as shown
in Fig.~2.  In spite of the spurious jitter, the two 
estimates agree to within 20\%, in the wide
quantum well limit.  We can also compare Eq.~(\ref{eq:final}) to the estimate 
$\Lambda \simeq (7.2\times 10^{-11})V_0$ obtained in Ref.~[\onlinecite{friesen07}], by 
fitting theoretical EM predictions to TB numerical solutions for the valley splitting.  
The latter  
differs from Eq.~(\ref{eq:final}) by about a factor of two, which we attribute to the 
fundamental limitations of the EM ansatz of Eqs.~(\ref{eq:TBF})-(\ref{eq:TBFS}).  
Nevertheless, it is clear that the EM theory and the multiscale analysis, described here,
capture the essential physics of valley splitting, and enable semi-quantitative predictions.
Thus justified, the more accurate numerical estimate for $\Lambda$ in 
Ref.~[\onlinecite{friesen07}] forms the basis for a fully quantitative EM theory.
Further improvements in the EM ansatz and the kink analysis should lead to better
correspondence between the estimates for $\Lambda$.

We now turn to direct gap materials, such as GaAs.  Although a sharp confinement potential
does not cause valley coupling in this case, there are still atomic scale corrections 
to the EM theory.  As in the Si case, the corrections tend to be more significant for narrow 
quantum wells \cite{long}.  Foreman has also noted that the corrections can be treated 
within the EM theory by introducing a $\delta$-function at the interface \cite{foreman05}.  
Here, we apply the multiscale theory developed for Si to the GaAs quantum well, obtaining 
an analytical expression for the GaAs interface potential, $\Lambda$.  We also compare
the improved wavefunction solutions to those obtained from TB theory. 

The simplest TB theory for a single ($\Gamma$)-valley material involves just 
the nearest-neighbor tunneling parameter $t_1=-8\hbar^2/m^*a^2$, where $a=5.64$~\AA~is the
width of the GaAs cubic unit cell.  For Al$_x$Ga$_{1-x}$As, used in the barriers, the effective 
mass $m^*$ depends on the composition $x$ to a greater degree than silicon alloys.  
Hence, $t_1$ depends on the atomic position.  The onsite parameter 
$\epsilon(z)=16\hbar^2/m^*a^2+V(z)$ also depends on composition.  However, for
the sake of transparency, we will ignore effective mass variations, taking $\epsilon(z)=V(z)$, 
and setting $m^*$ to an appropriate average of the effective masses near the
interface.  Indeed, $m^*$ eventually drops out of the leading order expression for $\Lambda$, 
and a more careful treatment provides only small corrections.

The matching condition, Eq.~(\ref{eq:BC}), also hold for GaAs.  
However, because there is only one valley, and the TB model has only one band,
the TB envelope function and wavefunction are now identical:  $C_j=F_j$.  
Near the interface, we parametrize the envelope as shown in the inset of Fig.~1.  
The TB eigenstates are smooth (in contrast with Fig.~2), so we may now use 
adjacent TB equations:
\begin{equation}
\begin{array}{l}
t_1C_{N-1}+t_1C_{N+1}=E_\text{TB}C_N ,\\
t_1C_N+V_0C_{N+1}+t_1C_{N+2}=E_{TB}C_{N+1} .
\end{array}
\end{equation}
In this case, there is no need to consider curvature of the wavefunction.  Solving for 
$\beta$ and $\gamma$ directly, and noting the relation $E_\text{TB}=2t_1+E$ between the 
TB and EM energies, we obtain the GaAs result,
\begin{equation}
\Lambda=a(V_0-E)/4 . \label{eq:GaAsmain}
\end{equation}
In the physically relevant limit of $E\ll V_0$, the GaAs and Si interface potentials
have identical forms.
For GaAs materials parameters, we obtain $\Lambda \simeq (1.41\times 10^{-10})V_0$.

The effect of the interface potential on the wavefunction is shown
in Fig.~1, where we plot the differences between the approximate 
(multiscale) and exact (TB) results for a narrow quantum well.  We also show results
for the conventional ($\Lambda =0$) EM theory.  Deviations from the EM theory are small 
in both cases.  However because there is no jitter in the GaAs TB envelope function, we 
find that Eq.~(\ref{eq:GaAsmain}) captures the atomic scale corrections with great accuracy.

In conclusion, we have demonstrated that leading corrections to the effective mass theory
at a sharp quantum well boundary arise from the atomic scale physics near the interface.  The
corrections appear as a small kink in the envelope function.
A multiscale approach, combining effective mass and tight binding theories, leads to 
an analytical expression for the effective interface potential in silicon, which determines 
the valley splitting.  Similar 
corrections apply to GaAs quantum wells, although there is no valley coupling.
For other device geometries, including graded interfaces and electric fields,  
the present approach remains robust.
These situations may also be treated by a multiscale analysis.

This work was supported by NSA/ARO contract no.\ W911NF-04-1-0389 and by NSF grant nos.\
DMR-0325634, CCF-0523675, and CCF-0523680.


\begin{thebibliography}{99}

\bibitem{zutic04}
I. Zutic, J. Fabian, and S. Das Sarma, Rev. Mod. Phys. \textbf{76}, 323 (2004).

\bibitem{loss98}
D. Loss and D. P. DiVincenzo, \pra \textbf{57}, 120 (1998).

\bibitem{kane98}
B. E. Kane, Nature (London) \textbf{393}, 133 (1998).

\bibitem{ando82}
T. Ando, A. B. Fowler, and F. Stern, Rev. Mod. Phys. \textbf{54}, 437 (1982).

\bibitem{goswami07}
S. Goswami, K. A. Slinker, M. Friesen, L. M. McGuire, J. L. Truitt, C. Tahan, L. J. Klein, 
J. O. Chu, P. M. Mooney, D. W. van der Weide, R. Joynt, S. N. Coppersmith, and M. A. Eriksson, 
Nat. Phys. \textbf{3}, 41 (2007). 

\bibitem{friesen03}
M. Friesen, P. Rugheimer, D. E. Savage, M. G. Lagally, 
D. W. van der Weide, R. Joynt, and M. A. Eriksson, \prb \textbf{67},
121301(R) (2003).

\bibitem{smelyanskiy05}
V. N. Smelyanskiy, A. G. Petukhov, and V. V. Osipov, \prb {72}, 081304(R) (2005).

\bibitem{kohn}
W.~Kohn, in \textit{Solid State Physics}, edited by F.~Seitz and 
D.~Turnbull (Academic Press, New York, 1957), Vol.~5.

\bibitem{ohkawa77}
F. J. Ohkawa and Y. Uemura, J. Phys. Soc. Japan \textbf{43}, 907 (1977);
\textit{ibid}, \textbf{43}, 917 (1977).

\bibitem{sham79}
L. J. Sham and M. Nakayama, \prb \textbf{20}, 734 (1979).

\bibitem{friesen05}
M. Friesen, \prl \textbf{94}, 186403 (2005).

\bibitem{ahn05}
D. Ahn, Journ. Appl. Phys. \textbf{98}, 033709 (2005).

\bibitem{friesen07}
M. Friesen, S. Chutia, C. Tahan, and S. N. Coppersmith,
\prb \textbf{75}, 115318 (2007) .

\bibitem{boykin04}
T. B. Boykin, G. Klimeck, M. A. Eriksson, M. Friesen, S. N. Coppersmith,
P. von Allmen, F. Oyafuso, and S. Lee, Appl. Phys. Lett. \textbf{84}, 115 (2004).

\bibitem{boykin04b}
T. B. Boykin, G. Klimeck, M. Friesen, S. N. Coppersmith,
P. von Allmen, F. Oyafuso, and S. Lee, \prb \textbf{70}, 165325 (2004).

\bibitem{burt92}
M. G. Burt, J. Phys. Cond. Mat. \textbf{4}, 6651 (1992);
M. G. Burt, \prb \textbf{50}, 7518 (1994).

\bibitem{foreman05}
B. A. Foreman, \prb \textbf{72}, 165345 (2005).

\bibitem{nestoklon06}
M.O. Nestoklon, L.E. Golub, and E.L. Ivchenko, 
\prb \textbf{73}, 235334 (2006).

\bibitem{fritzsche}
H. Fritzsche, Phys. Rev. \textbf{125}, 1560 (1962).

\bibitem{twose}
W. D. Twose, in the Appendix of Ref.~[\onlinecite{fritzsche}].

\bibitem{daviesbook}
J. H. Davies, \textit{The Physics of Low-Dimensional Semiconductors}
(Cambridge Press, Cambridge, 1998).

\bibitem{long}
F. Long, W. E. Hagston, and P. Harrison, in \textit{23rd International
Conference on the Physics of Semiconductors}, edited by M.
Sheffler and R. Zimmermann (World Scientific, Singapore,
1996) p. 819.

\end{thebibliography}
\end{document}